\documentclass[preprint,superscriptaddress,amsmath,amssymb]{revtex4}

\usepackage{graphicx}
\usepackage{dcolumn}
\usepackage{bm}
\usepackage{ulem}
\usepackage{epstopdf}
\usepackage{amsmath}
\usepackage{amssymb}

\newcommand{\nc}{\newcommand}
\nc{\be}{\begin{eqnarray}}
\nc{\ee}{\end{eqnarray}}
\nc{\bea}{\begin{eqnarray}}
\nc{\eea}{\end{eqnarray}}
\nc{\bean}{\begin{eqnarray*}}
\nc{\eean}{\end{eqnarray*}}

\preprint{}

\begin{document}

\title{In-plane electronic anisotropy in the antiferromagnetic-orthorhombic phase of isovalent-substituted Ba(Fe$_{1-x}$Ru$_x$)$_2$As$_2$}

\author{L.~Liu}
\affiliation{Department of Physics, University of Tokyo, Tokyo 113-0033, Japan}

\author{T.~Mikami}
\affiliation{Department of Physics, University of Tokyo, Tokyo 113-0033, Japan}

\author{S.~Ishida}
\affiliation{National Institute of Advanced Industrial Science and Technology, Tsukuba 305-8568, Japan}

\author{K.~Koshiishi}
\affiliation{Department of Physics, University of Tokyo, Tokyo 113-0033, Japan}

\author{K.~Okazaki}
\affiliation{Department of Physics, University of Tokyo, Tokyo 113-0033, Japan}

\author{T.~Yoshida}
\affiliation{Graduate school of human and environmental studies, Kyoto University, Kyoto 606-8502, Japan}

\author{H.~Suzuki}
\affiliation{Department of Physics, University of Tokyo, Tokyo 113-0033, Japan}

\author{M.~Horio}
\affiliation{Department of Physics, University of Tokyo, Tokyo 113-0033, Japan}

\author{L.~C.~C.~Ambolode~II}
\affiliation{Department of Physics, University of Tokyo, Tokyo 113-0033, Japan}

\author{J.~Xu}
\affiliation{Department of Physics, University of Tokyo, Tokyo 113-0033, Japan}

\author{H.~Kumigashira}
\affiliation{KEK, Photon Factory, Tsukuba, Ibaraki 305-0801, Japan}

\author{K.~Ono}
\affiliation{KEK, Photon Factory, Tsukuba, Ibaraki 305-0801, Japan}

\author{M.~Nakajima}
\affiliation{National Institute of Advanced Industrial Science and Technology, Tsukuba 305-8568, Japan}

\author{K.~Kihou}
\affiliation{National Institute of Advanced Industrial Science and Technology, Tsukuba 305-8568, Japan}

\author{C.~H.~Lee}
\affiliation{National Institute of Advanced Industrial Science and Technology, Tsukuba 305-8568, Japan}

\author{A.~Iyo}
\affiliation{National Institute of Advanced Industrial Science and Technology, Tsukuba 305-8568, Japan}

\author{H.~Eisaki}
\affiliation{National Institute of Advanced Industrial Science and Technology, Tsukuba 305-8568, Japan}

\author{T.~Kakeshita}
\affiliation{Department of Physics, University of Tokyo, Tokyo 113-0033, Japan}

\author{S.~Uchida}
\affiliation{Department of Physics, University of Tokyo, Tokyo 113-0033, Japan}

\author{A.~Fujimori}
\affiliation{Department of Physics, University of Tokyo, Tokyo 113-0033, Japan}




\begin{abstract}

 We have studied the anisotropy in the in-plane resistivity and the electronic structure of isovalent Ru-substituted BaFe$_2$As$_2$ in the antiferromagnetic-orthorhombic phase using well-annealed crystals. The anisotropy in the residual resistivity component increases in proportional to the Ru dopant concentration, as in the case of Co-doped compounds. On the other hand, both the residual resistivity and the resistivity anisotropy induced by isovalent Ru substitution is found to be one order of magnitude smaller than those induced by heterovalent Co substitution. Combined with angle-resolved photoemission spectroscopy results, which show almost the same anisotropic band structure both for the parent and Ru-substituted compounds, we confirm the scenario that the anisotropy in the residual resistivity arises from anisotropic impurity scattering in the magneto-structurally ordered phase rather than directly from the anisotropic band structure of that phase.

\end{abstract}

\maketitle


\section{introduction}

A typical parent compound of iron-based superconductors, e.g., BaFe$_2$As$_2$ (Ba122), shows a tetragonal-orthorhombic structural phase transition as well as a paramagnetic (PM)-antiferromagnetic (AFM) transition at the same temperature. Either by chemical doping or applying external pressure, the non-superconducting AFM parent compound is driven to a superconducting (SC) state. The phase diagrams of such chemically doped iron pnictides are similar to those of several other classes of unconventional superconductors, including the cuprates and heavy-fermion superconductors, in that they all show a SC dome in close proximity to a phase with magnetism~\cite{Paglione_NP10}. It is, therefore, believed that a comprehensive understanding of the metallic AFM ground state in the parent compounds of iron-based superconductors is of primary importance for elucidating the origin of the high-temperature superconductivity. In the most extensively studied 122-type iron pnictides, there are mounting evidences suggesting that the antiferromagnetic-orthorhombic (AFO) phase is unique with a hallmark of the electronic anisotropy~\cite{Fisher_RPP11}. Early resistivity measurements on detwinned Ba122 crystals revealed strong resistivity anisotropy in the AFO phase. Unexpectedly, the resistivity along the $b$ axis [shorter axis with ferromagnetic (FM) spin alignment] is larger than that along the $a$ axis (longer axis with AFM spin alignment)~\cite{Chu_science10}. The origin of the resistivity anisotropy has been discussed in terms of spin-fluctuation scattering~\cite{Fernandes_PRL11, Blomberg_NC13}, anisotropic reconstructed Fermi surface (FS)~\cite{Fisher_RPP11, Kuo_PRB11}, and orbital ordering~\cite{Chen_PRB10}. On the other hand, recent investigations of in-plane resistivity anisotropy on well-annealed Ba(Fe$_{1-x}$Co$_x$)$_2$As$_2$ (Co-Ba122) crystals demonstrated that the in-plane resistivity anisotropy arises from the elastic impurity scattering by doped Co atoms~\cite{Ishida_PRL13, Nakajima_PRL12}. A scanning tunneling microscopy (STM) study has revealed that an anomalous, anisotropic impurity state around a dopant atom is formed, which is proposed to account for the sizable in-plane resistivity anisotropy~\cite{Allan_NP13}. This exotic impurity state around the doped Co atoms needs to be further investigated for the systems with different kinds of doping/impurities.

\indent Angle-resolved photoemission spectroscopy (ARPES) measurements on Co-Ba122 in the AFO phase clarified an anisotropic feature of the FS and the $d_{xz}$/$d_{yz}$ band splitting near the corner of the two-dimensional Brillouin zone (BZ), reflecting the broken four-fold rotational symmetry~\cite{Yi_PNAS11}. It was shown by the ARPES measurement that the FS consists of two hole pockets and one electron pocket centered at the BZ center, surrounded by two bright spots along the $\Gamma$-X line and two bigger petal pockets along the $\Gamma$-Y  line~\cite{Yi_PNAS11}. The relationship between the resistivity anisotropy observed by transport measurements and the anisotropic electronic structure probed by ARPES measurements remains elusive.

In Ba(Fe$_{1-x}$Ru$_x$)$_2$As$_2$ (Ru-Ba122), Ru substitutes for Fe atoms as in the case of Co-Ba122, while its substitution is nominally isovalent and would not change the band filling. On the other hand, Ru is a 4$d$ element and is chemically quite different from the 3$d$ element. It is interesting to study the effect of Ru substitution on the resistivity anisotropy and compare it with that of the Co-doped system. It should be noted that as-grown Ba122 crystals tend to have defects in the FeAs blocks which are most likely responsible for the observed finite resistivity anisotropy~\cite{Ishida_PRB11}, confusing the study of the genuine effect of the dopants. The annealing process with BaAs powders has been demonstrated to be an effective method to reduce the crystal defects, evidenced by the remarkable reduction in the residual resistivity (RR) and an increase in the magneto-structural phase transition temperature~\cite{Ishida_PRB11, Nakajima_PNAS11}. Thus, measurements on annealed crystals with reduced defects are indispensable for the study of the resistivity anisotropy. 

In this paper, we first present an investigation of the in-plane resistivity of well-annealed Ru-Ba122 crystals. We shall show that the resistivity anisotropy induced by Ru substitution emerges in the AFO phase, having the same sign as that in Co-Ba122, while the magnitude in Ru-Ba122 is much smaller. Furthermore, we shall demonstrate that the anisotropic electronic structures in the tetragonal-symmetry-broken AFO phase remains almost unchanged with slight Ru substitution by ARPES measurements using detwinned crystals, confirming that the resistivity anisotropy results from the impurity effect in the anisotropic ground state.


\section{experimental methods}

Single crystals of Ba(Fe$_{1-x}$Ru$_x$)$_2$As$_2$ (nominal Ru content $x$ = 0, 0.08, 0.15) were grown by the self-flux method and post-annealed as described in Refs.~\cite{Ishida_PRL13, Ishida_PRB11, Nakajima_PNAS11}. The Ru compositions of the samples were determined by energy-dispersive X-ray spectroscopy (EDXS) analysis. It was found that the actual Ru contents of the doped samples were $x$ = 0.04 and 0.08, respectively, almost half of the nominal ones. Relationship between the actual content and the nominal one in the underdoped region is very similar to that reported in Ref.~\cite{Thaler_PRB10}. Hereafter, the actual $x$ values are used in this paper. The in-plane resistivity of the twinned samples was measured using a standard four-terminal method. For the measurement of in-plane resistivity anisotropy, the crystals were detwinned by applying uniaxial compressive pressure and the resistivity along the $a$ and $b$ axes were measured using the Montgomery method as described elsewhere~\cite{Ishida_PRB11, Liang_JPCS11, Montgomery_JAP71}. The measurements were performed in a Quantum Design Physical Property Measurement System (PPMS) and all the data were obtained while warming the samples.

For the investigation of the electronic structure, we performed ARPES measurements at beamline BL-28A of Photon Factory (PF) using a SCIENTA SES2002 electron analyzer. In order to measure the intrinsic in-plane electronic anisotropy hampered by the twin formation in the AFO phase, we developed a mechanical detwinning device similar to that used in Ref. ~\cite{Kim_PRB11} for ARPES measurements. Circularly-polarized light was used with the photon energy of 63 eV, corresponding to $k_z$ $\sim$ 2$\pi$/c. The total energy resolution was set to $\sim$ 20 meV. The crystals were cleaved \textit{in situ} at $T$ = 20 K and measured in an ultra-high vacuum of ~ $1\times10^{-10}$ Torr. Calibration of the Fermi level ($E_F$) was achieved by referring to that of gold.


\section{Results and Discussion}


Figure 1 presents the temperature dependence of the in-plane resistivity for underdoped Ru-Ba122 crystals. The structural transition temperature ($T_s$) and magnetic transition temperature ($T_N$) determined from the derivative of the resistivity curves using the method of Ref.~\cite{Chu_PRB09} do not split as in the case of Co-Ba122. $T_s \approx T_N$ systematically decreases upon Ru substitution, 142 K for $x$ = 0, 128 K for $x$ = 0.04, and 116 K for $x$ = 0.08. The resistivity in the PM state also decreases with Ru substitution, as in the case of Co-, P-doped Ba122~\cite{Ishida_JACS13}. It has been theoretically demonstrated for Ru-Ba122 that the electronic states near the chemical potential is insensitive to the randomly distributed Ru impurity scattering, resulting from a coherent interference of correlated on-site and intersite impurity scattering~\cite{Wang_PRL13}. The residual resistivity of undoped Ba122 is quite small ($\sim$ 10 $\mu\Omega$ cm) for the well-annealed crystal and it becomes larger upon Ru substitution as seen in Fig. 1(a). The increase of RR can be attributed to the elastic scattering of doped Ru atoms, evidenced by the linear increase of RR with the Ru content $x$. The recent study of the in-plane resistivity of BaFe$_2$As$_2$ doped at three different lattice sites (K for Ba, Co for Fe, and P for As) has shown that the impurity scattering by dopant atoms in the AFO phase becomes weaker as the dopant sites move away from the Fe plane~\cite{Ishida_JACS13}. However, compared with Co doping into the Fe plane, Ru doping into the Fe plane turns out to exhibit a much weaker effect on RR. The results for the annealed Co-, Ru-, and P-substituted BaFe$_2$As$_2$ single crystals are summarized in Fig. 2(a). One can see that the doping-induced RR for the well-annealed Ru- and P-substituted Ba122 is just on the same line. Ru substitution for Fe is thought to be another way to realize isovalent substitution besides P substitution for As~\cite{Xu_PRB12}. The present results demonstrate that atoms which are isovalently substituted into the FeAs blocks have a surprisingly similar effect on the scattering of carriers.

Figure 1(b) shows the in-plane resistivity anisotropy measured on detwinned Ru-Ba122 crystals. The anisotropy, especially in the RR component, vanishes for undoped compounds, while it gradually increases upon Ru substitution in the AFO phase. The anisotropy of Ru-Ba122 is found to have the same sign as that of Co-Ba122: Carriers move more easily along the longer and AFM spin-aligned $a$ axis than the shorter and FM spin-aligned $b$ axis. The resistivity anisotropy of the Ru-doped samples persists well above $T_s \approx T_N$. Perhaps this is related to the broadening of the first-order magneto-structural transition due to the disorder caused by Ru substitution.
Here, we focus on the resistivity anisotropy in the temperature-independent RR component. Similar to the case of Co-Ba122~\cite{Ishida_PRL13}, a linear increase in the magnitude of the RR anisotropy was clearly found with the increase in the Ru content $x$, indicating that the impurity-induced-anisotropy mechanism proposed for Co-Ba122 also works for Ru-Ba122. We have plotted the magnitude of the resistivity anisotropy at 5 K against the dopant content for Co-, Ru- and P-substitution in Fig. 2(b). For BaFe$_2$(As$_{1-x}$P$_x$)$_2$ (P-Ba122), measurements of the resistivity anisotropy has been reported only on as-grown samples so far, from which the genuine resistivity anisotropy caused by P atoms was estimated~\cite{Ishida_JACS13}. It is intriguing that the isovalent Ru- and P-substitution have the same effect on both RR and resistivity anisotropy, regardless of the substitution sites, namely, at the Fe sites or the As sites. This implies that the As layer plays an equally important role in building the conducting layers in the iron arsenide. From the comparison between Ru and Co substitutions, the magnitude of anisotropy of Ru-Ba122 is almost one order smaller, indicating that the heterovalent Co substitution has a much stronger impurity effect on inducing the resistivity anisotropy than the isovalent Ru substitution. Both the RR and the anisotropy induced by the same number of Co atoms are about one order of magnitude larger than that of isovalent dopant Ru or P.

To summarize the transport results, we observed the emergence of the in-plane resistivity anisotropy in the AFO phase upon Ru substitution in the Ba122 system. The resistivity along the shorter and FM axis ($b$ axis) becomes larger, similar to that observed in 3$d$ transitional metal-doped systems~\cite{Kuo_PRB11, Ishida_PRL13}. Compared with Co doping, while the anisotropy of Ru-Ba122 is found to be much smaller, its origin can also be attributed to the anisotropic scattering effect of the Ru impurity as has been proposed for Co-Ba122.

Figure 3 shows ARPES results on the FS in the AFO phase of undoped and Ru-doped Ba122 measured on detwinned crystals. The three-dimensional (3D) BZ of the PM phase is folded into the smaller 3D BZ in the AFO phase as illustrated in Fig. 3(a). 63 eV photons were used to probe the electronic structures at $k_z$ $\sim$ 2$\pi$/c. In Fig. 3(b), clear anisotropy in the FS mapping intensity of  BaFe$_2$As$_2$ along the $a$ and $b$ axes can be seen. In order to see whether this anisotropy is intrinsic or due to matrix-element effect, we also performed the measurements with the sample rotated in the $k_x-k_y$ plane by $90^{\circ}$ as shown in Fig. 3(c). In the figure, the ARPES intensity mapping is rotated by $90^{\circ}$ in the $k_x-k_y$ plane, indicating that the observed anisotropy in the FS mapping is intrinsic and that the samples in our measurements were effectively detwinned. The anisotropic FS topology reflects the broken fourfold rotational (tetragonal) symmetry in the AFO state. From the present ARPES measurements, we resolved the FS in the AFO phase consisting of three types of pockets: Isotropic hole pocket centered at the BZ center (denoted as $\alpha$), tiny rounded electron pocket along the Z-X line originating from the Dirac-cone-like band crossing (denoted as $\gamma$), and larger elliptical electron pocket along the Z-Y line arising from the band reconstruction (denoted as $\delta$). A schematic figure of the resolved FS pockets in the folded BZ is illustrated in the inset of Fig. 3(b). Figure 3(d) presents the FS mapping intensity of Ba(Fe$_{0.96}$Ru$_{0.04}$)$_2$As$_2$ with the same measurement geometry as Fig. 3(b). No discernible change was observed in the anisotropy of FS with Ru doping. We also note that among the resolved FS pockets in-plane mass anisotropy can only be identified in electron pocket $\delta$, where the effective mass $m^*$ along the $a$ axis is larger than along the $b$ axis. Thus, the mass anisotropy cannot account for the observed resistivity anisotropy that the resistivity along the $a$ axis is smaller.

Figure 4(a)-(d) shows band dispersions along the Z-X and Z-Y high symmetry lines of BaFe$_2$As$_2$ and Ba(Fe$_{0.96}$Ru$_{0.04}$)$_2$As$_2$. The $d_{yz}$ band around the X point and the $d_{xz}$ band around the Y point become inequivalent as marked by red broken lines, manifesting the broken fourfold rotational symmetry in the electronic structure. The same difference between the $d_{yz}$ band along the Z-X line and the $d_{xz}$ band along the Z-Y line is seen for Ba(Fe$_{0.96}$Ru$_{0.04}$)$_2$As$_2$. These results evidence that the emergence of the resistivity anisotropy does not arise from the anisotropic electronic structure caused by orthorhombic lattice distortion but arises from the extrinsic impurity effect of dopants.

The resistivity anisotropy of the Ru-doped samples persists well above $T_s/T_N$. Fernandes \textit{et al.} showed that the anisotropic resistivity could be induced by an anisotropic dressing of the impurity scattering by spin fluctuations~\cite{Fernandes_PRL11}. Alternatively, this might arise from a broadening of the first-order magneto-structural transition due to the Ru disorder and/or from the local strain produced by Ru with different ionic radius from that of Fe which might stabilize local AFO order around it. In order to further clarify the interplay between impurities and the nematic electronic state above $T_s/T_N$, detailed experimental study of the FS and the quasiparticle scattering rates in the paramagnetic-tetragonal phase is highly needed.


\section{conclusion}

We investigated the in-plane electronic anisotropy of isovalent-substituted Ru-Ba122 by transport measurements and ARPES measurements on well-annealed crystals. The in-plane resistivity anisotropy in the RR component was found to emerge upon Ru substitution. Similar to that in Co-Ba122, the resistivity along the shorter and FM axis, $\rho_b$, has a larger magnitude than $\rho_a$ and the magnitude of the anisotropy in the RR component linearly increases with the Ru content. Moreover, it was found that both the magnitude of the resistivity and the resistivity anisotropy induced by isovalent Ru atoms is one order of magnitude smaller than that induced by heterovalent Co substitution, while they are almost identical to that induced by P atoms which isovalently substitute for the As sites. Our ARPES measurements on detwinned crystals demonstrated the FS topology in the AFO phase: Isotropic hole pocket, Dirac-cone-like tiny electron pocket along the Z-X line and elliptical electron pocket with much larger anisotropy along the Z-Y line. Upon slight Ru substitution, no discernible change in the anisotropic FS or the $d_{xz}$/$d_{yz}$ band splitting around the BZ corner was resolved. Our results demonstrate the extrinsic nature of the resistivity anisotropy in the temperature-independent RR component: it originates from the anisotropic scattering of doped atoms in the AFO phase.


\section*{Acknowledgements} 

We thank D. Hirai for his help in the transport measurement. Transport measurement was also performed using facilities at the Cryogenic Research Center, the University of Tokyo. This work was supported by the Japan-China-Korea A3 Foresight Program from the Japan Society for the Promotion of Science (JSPS). Experiment at Photon Factory was approved by the Photon Factory Program Advisory Committee (Proposal No. 2012S2-001 and 2012G075). L.L. thanks the Ministry of Education, Culture, Sports, Science, and Technology (MEXT) Scholarship program of Japan and China Scholarship Council (CSC) for financial support.

\newpage


\newpage

\begin{figure}[htb]
\begin{center}
\includegraphics[width=16cm]{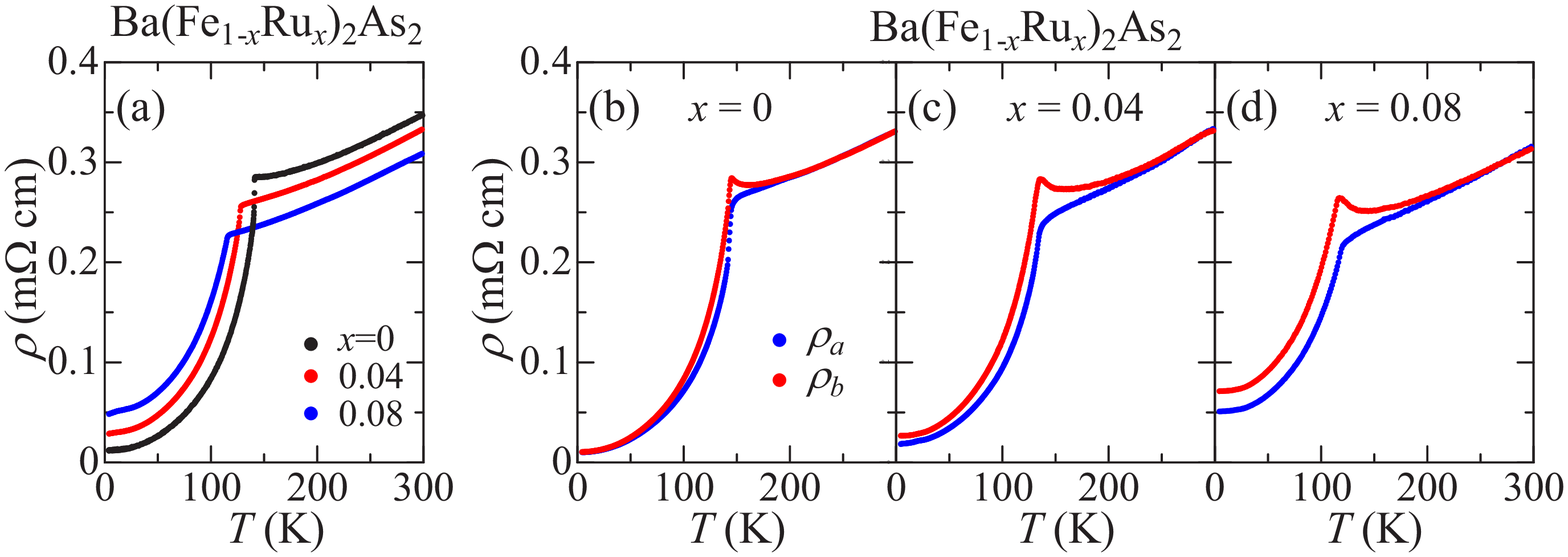}
\caption{\label{fig1} (color online) Temperature dependence of the in-plane resistivity of Ba(Fe$_{1-x}$Ru$_x$)$_2$As$_2$ crystals with $x$ = 0, 0.04, and 0.08. (a) Resistivity measured on twinned crystals. (b)-(d) $\rho_a$ and $\rho_b$ measured on detwinned crystals.}
\end{center}
\end{figure}

\begin{figure}[htb]
\begin{center}
\includegraphics[width=7.5cm]{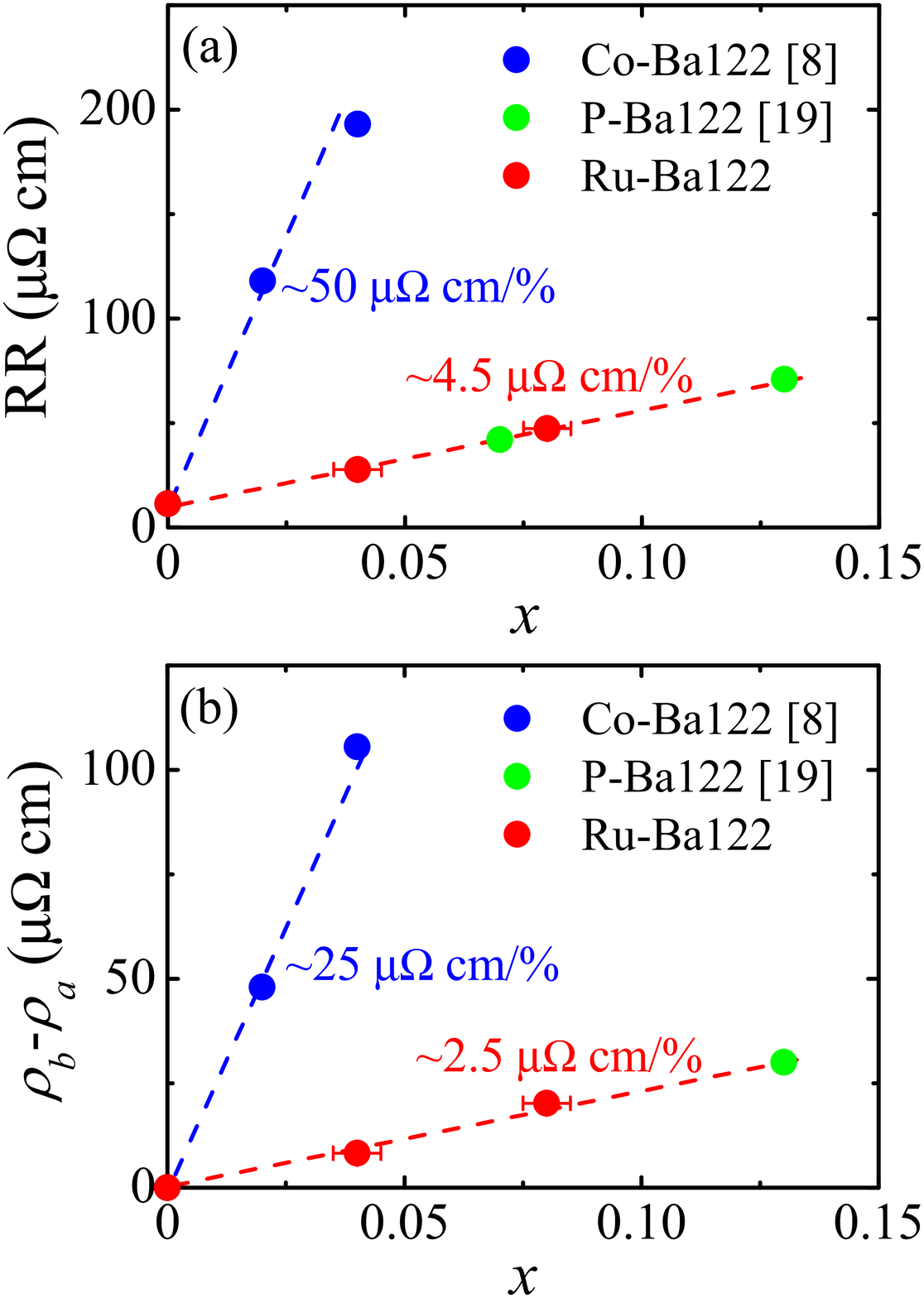}
\caption{\label{fig2} (color online) Residual resistivity (a) and in-plane resistivity anisotropy ($\rho_b - \rho_a$) (b) at 5 K plotted against the dopant content $x$ for doped BaFe$_2$As$_2$ crystals with three types of substitution: isovalent Ru for Fe; heterovalent Co for Fe~\cite{Ishida_PRL13}; isovalent P for As~\cite{Ishida_JACS13}. }
\end{center}
\end{figure}

\begin{figure}[htb]
\begin{center}
\includegraphics[width=16cm]{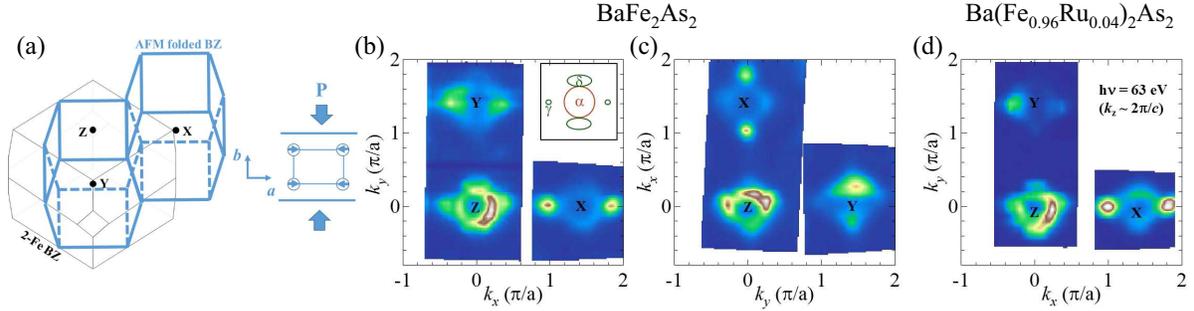}
\caption{\label{fig3} (color online) (a) 3D BZ of the bct structure for the PM state (grey) and the folded BZ for the AFO state (blue). Notation in the PM BZ is used, where Z-X is along the AFM direction and Z-Y is along the FM direction. The right figure illustrates the crystal axes in the AFO state after being detwinned by applying uniaxial compressive pressure. (b) FS mapping in the $k_x-k_y$ plane of detwinned BaFe$_2$As$_2$ measured at 20 K. All FS mappings in this paper were made within an integration window of $E_F$ $\pm$ 10 meV. A schematic figure of the resolved FS is illustrated in the inset of (b). (c) The same as (b) with the samples rotated in the $k_x-k_y$ plane by $90^{\circ}$. (d) The same FS mapping as (b) for Ba(Fe$_{0.96}$Ru$_{0.04}$)$_2$As$_2$.}
\end{center}
\end{figure}

\begin{figure}[htb]
\begin{center}
\includegraphics[width=10cm]{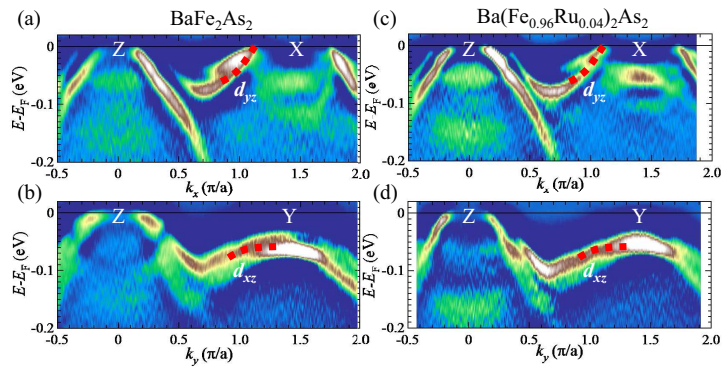}
\caption{\label{fig4} (color online) (a)-(b) Band structures of detwinned BaFe$_2$As$_2$ crystals along the high symmetry lines Z-X and Z-Y. Spectral images have been obtained through energy second-derivatives of the raw images. Red broken lines are drawn as a guide to the eye to illustrate the anisotropic band dispersions around the X/Y point. (c)-(d) The same spectral images for detwinned Ba(Fe$_{0.96}$Ru$_{0.04}$)$_2$As$_2$ crystals.}
\end{center}
\end{figure}

\end{document}